\documentclass[final,1p,times]{elsarticle}

\usepackage{graphicx}
\usepackage{amssymb}
\usepackage{amsthm}
\usepackage{lineno}

\journal{Nuclear Physics A}

\begin{document}

\begin{frontmatter}

\title{Jet probes of QCD matter: \\single jets and dijets in heavy-ion collisions}



\author[auth1,auth2]{Ben-Wei Zhang}
\author[auth1]{Yuncun He}
\author[auth1]{Enke Wang}
\address[auth1]{Key Laboratory of Quark and Lepton Physics (MOE) and Institute of Particle Physics, \\
 Central China Normal University, Wuhan 430079, China}
\address[auth2]{Nuclear Science Division, MS 70R0319, Lawrence Berkeley National Laboratory, Berkeley, CA 94720}

\begin{abstract}
Modifications of jets in the existence of a hot and dense QCD medium have recently attracted a lot of attentions. In this talk, we demonstrate how jet-medium interactions change the behavior of jets by offering examples of inclusive jet and dijet productions at $ {\cal O}( \alpha_s^3 )$  in heavy ion collisions 
including initial-state cold nuclear effects and especially the final-state parton energy loss effect. The suppression of inclusive jet spectrum varying with jet radii 
and a flatter dijet momentum imbalance as compared to those in hadron-hadron collisions are observed in high-energy nuclear collisions.
\end{abstract}

\end{frontmatter} 


\section{Introduction}

An Energetic parton traversing the QCD matter should lose energy due to its inelastic and elastic scattering with other partons 
in the medium~\cite{Wang:1991xy, Baier:1996sk, Gyulassy:2000fs,Wang:2001ifa,Neufeld:2011yh}.  This attenuation of parton energy will 
leave its fingerprint not only on leading hadron spectra but also on jet productions  in relativistic heavy-ion collisions. 
 In the following we will discuss single jets and dijets productions in high energy heavy-ion reactions to show how the jet-medium interactions in the QCD matter modify jet properties~\cite{Vitev:2008rz,Vitev:2009rd,He:2011sg,He:2011pd}. The same approach has also been used to investigate electroweak boson associated jet productions in heavy-ion collisions~\cite{ew-boson}.

\section{Inclusive jet productions in HIC}

With the jet differential cross section in p+p collisions as the input, we can obtain the medium-modified inclusive jet cross section with final-state
parton energy loss effect~\cite{Vitev:2008rz,Vitev:2009rd}:
\begin{eqnarray}
&&\frac{1}{\langle  N_{\rm bin}  \rangle}
 \frac{d\sigma^{AA}(R)}{dy_1dE_{T\,1}} =  \sum_{q,g}
\int_{\epsilon_1=0}^1 d\epsilon_1 \;  P_{q,g}(\epsilon_1,E_{T\,1})
 \times \frac{1}{ \left(1 - [1-f(R_1,p_{T\,1}^{\min})_{q,g}]  \epsilon_1\right)}
 \frac{d\sigma_{q,g}(E_{T\,1}^\prime)} {dy_1 dE^{\, \prime}_{T\,1}} \; . \qquad
\label{incl}
\end{eqnarray}
where $f(R_1,p_{T\,1}^{\min})_{q,g}$ is the part of the fractional energy loss that is redistributed inside the jet cone, and $p_{T\,1}^{\min}$ is a parameter which can be used to simulate processses responding for energy loss beyond bremsstrahlung. $P_{q,g}(\epsilon_1,E_{T\,1})$ is the probability that jet lose energy fraction $\epsilon_1$. $E^{\, \prime}_{T\,1}$ is related to $E_{T\,1}$ as $E^{\, \prime}_{T\,1}=E_{T\,1}/\left(1 - [1-f(R_1,p_{T\,1}^{\min})_{q,g}]  \epsilon_1\right)$.

\begin{figure}[htbp]
\begin{center}
 \includegraphics[width=0.45\textwidth]{LHCcouplingCH.eps}
 \includegraphics[width=0.45\textwidth,height=2.5in]{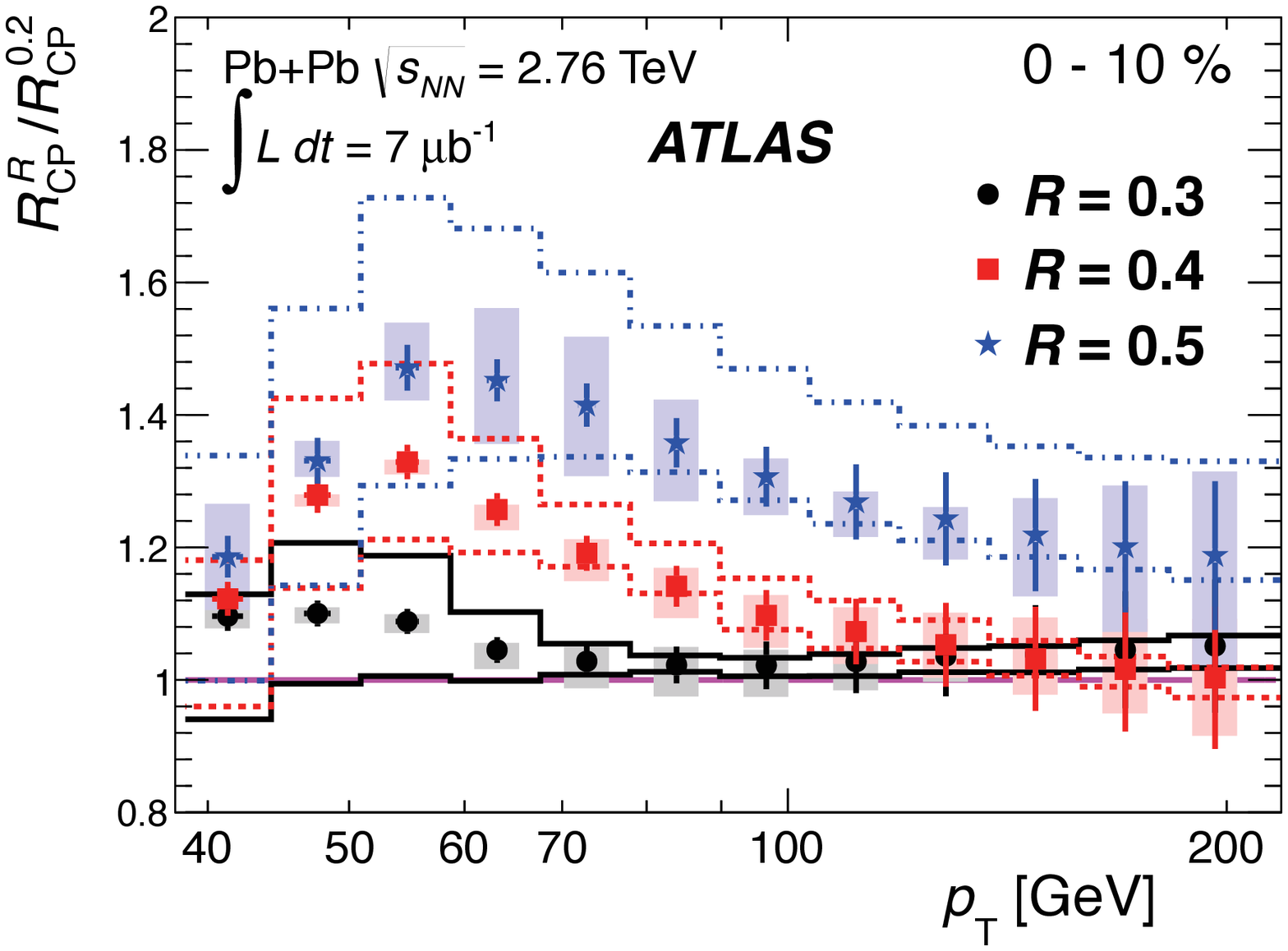}
\end{center}
\caption{The left panel: the dependence of suppression ratio for inclusive jet in central Pb+Pb collisions at $\sqrt{s}=2.76~$TeV on jet size. The right panel: ratios of $R_{CP}$ values between $R = 0.3, 0.4, 0.5$ and $R=0.2$ jets in central Pb+Pb collisions~\cite{:2012is}. }
\label{1}
\end{figure}

Fig.~\ref{1} displays the sensitivity of the suppression ratio for inclusive jet in central Pb+Pb collisions at $\sqrt{s}=2.76~$TeV to the jet size $R$. In the top plot of the left panel, it is the result involving both initial state and final state effects, and different styles of lines present different coupling strength of jet with the medium. 
It is illustrated that  the suppression is larger for the jet with the smaller size, and less pronounced with a larger jet size.  We see that the inclusive jet spectrum is suppressed further with the initial state cold nuclear matter effects. The predicted continuous variation of single jet suppression with the jet size 
has been observed recently by ATLAS collaboration~\cite{:2012is} as shown in the right panel of Fig.~\ref{1}.

\section{Dijet productions in HIC}

Dijet spectrum in heavy ion collisions is ready to be given similarly as the inclusive jet~\cite{He:2011pd},
\begin{eqnarray}
\frac{1}{\langle  N_{\rm bin}  \rangle}
\ \frac{d\sigma^{AA}(R)}{dy_1dy_2dE_{T\,1}dE_{T\,2} } &=&  \sum_{qq,qg,gg}
\int_{\epsilon_1=0}^1 d\epsilon_1 \int_{\epsilon_2=0}^1 d\epsilon_2
 \frac{P_{q,g}(\epsilon_1,E_{T\,1})}{ \left(1 - [1-f(R_1,p_{T\,1}^{\min})_{q,g}]
 \epsilon_1\right)}\nonumber \\
 &~&\frac{P_{q,g}(\epsilon_2,E_{T\,2})}{ \left(1 - [1-f(R_2,p_{T\,2}^{\min})_{q,g}]  \epsilon_2\right)}
 \times
 \frac{d\sigma_{qq,qg,gg}(E_{T\,1}^{\, \prime}, E_{T\,2}^{\,\prime})}
{dy_1 dy_2 dE^{\, \prime}_{T\,1} dE^{\,\prime}_{T\,2} } \; . \qquad
\label{eq:JCS-AA}
\end{eqnarray}
The two jets have symmetric final transverse momentum at leading-order. This symmetry is broken by  $2 \rightarrow 3$ parton splitting  processes at higher
corrections. To quantify the transverse momentum imbalance between the leading jets, we define the dijet transverse momentum
asymmetry,
\begin{eqnarray}
A_J=\frac{E_{T1}-E_{T2}}{E_{T1}+E_{T2}},
\label{eq:JCS-AA}
\end{eqnarray}
where $E_{T1}$ and $E_{T2}$ represent the transverse momentum for leading jet and subleading jet respectively. Thus the $A_J$ differential
distribution can be obtained through a change of variables,
\begin{eqnarray}
\frac{d\sigma}{dA_J}&=& \int_{E_{T\,2\, \min}}^{E_{T\,2\,\max}} dE_{T\, 2}
 \frac{2 E_{T\,2}}{(1-A_J)^2}\frac{d\sigma[E_{T\,1}(A_J,E_{T\,2})]}{dE_{T\,1}dE_{T\,2}} \;.
\label{ajcalc}
\end{eqnarray}
The transverse momentum fraction $z=E_{T2}/E_{T1}$ plays the same role with the asymmetry $A_J$. The $z$ differential distribution can be achieved,
\begin{eqnarray}
\frac{d\sigma}{dz}=\int_{E_{T\,2\, \min}}^{E_{T\,2\,\max}} dE_{T\, 2}  \; \frac{ E_{T\,2}}{z^2}
\times\frac{d\sigma[E_{T\,1}(z,E_{T\,2})]}{dE_{T\,1}dE_{T\,2}} \;.
\label{zcalc}
\end{eqnarray}

\begin{figure}[htbp]
\begin{center}
 \includegraphics[width=0.5\textwidth]{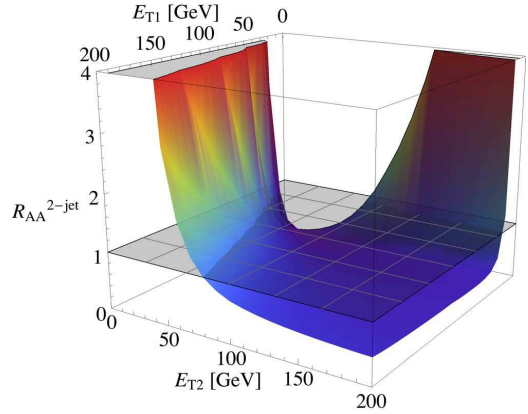}
\end{center}
\caption{The 3D suppression ratio of dijet cross sections in central Pb+Pb collisions at $\sqrt{s}=2.76~$TeV. The jet sizes are set $R_1=R_2=0.2$. }
\label{2}
\end{figure}

\begin{figure}[htbp]
\begin{center}
 \includegraphics[width=0.45\textwidth]{AJnewAAcoup.eps}
 \includegraphics[width=0.45\textwidth]{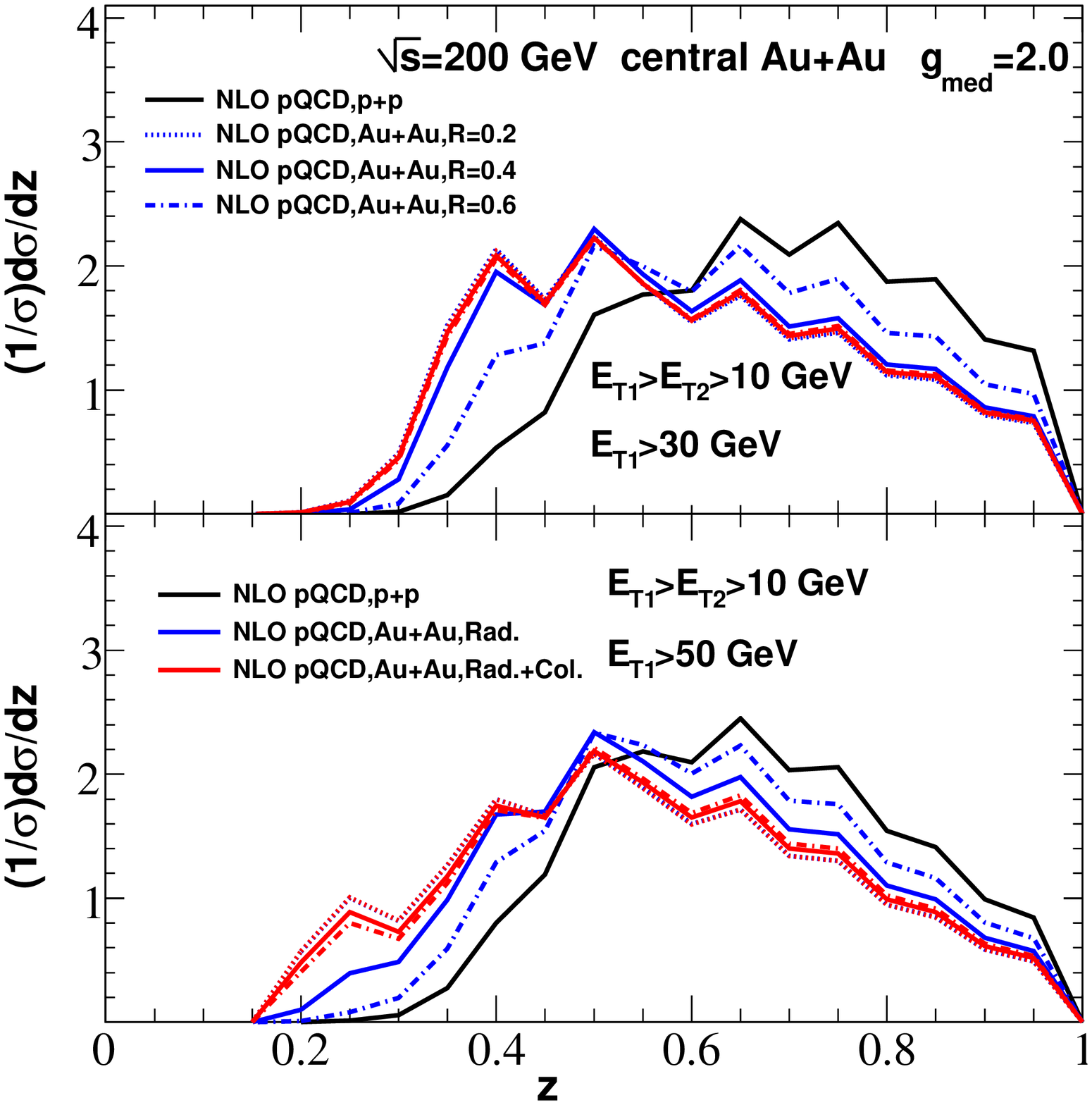}
\end{center}
\caption{The left panel: the normalized dijet $A_J$ distributions for different coupling strength in central Pb+Pb collisions at LHC. The right panel: the normalized dijet $z$ distributions for different radii in central Au+Au collisions at RHIC. }
\label{3}
\end{figure}
Fig.~\ref{2}~shows the suppression ratio of dijet at $ {\cal O}( \alpha_s^3 )$ in central Pb+Pb collisions at $\sqrt{s}=2.76~$TeV as the function of $E_{T1}$ and $E_{T2}$. The suppression is the largest along the diagonal regime where $E_{T1}=E_{T2}$. It varies slowly in the wide regime until the regime $E_{T1}$ far away from $E_{T2}$ where the ratio is strikingly enhanced. The medium effects dissipate partial energy of jet so that jet yields are redistributed to smaller transverse energies. These would result in the features of dijet momentum imbalance distributions in heavy ion reactions.

Fig.~\ref{3}~illustrates the normalized dijet momentum imbalance distributions in heavy ion reactions, and the black lines in the boxes present relevant results in p+p collisions. The left panel shows the dependence of $A_J$ distributions in central Pb+Pb collisions at $\sqrt{s}=2.76~$TeV, as an example of the size $R=0.4$, on coupling strength of jet to the medium. The distribution in  Pb+Pb collisions is broadened relative to that in  p+p collisions, due to the energy loss of jet induced by QGP medium. In the case of $p_{T}^{\min}=0~$GeV, the broadening does not reach the extent  observed by experiments yet, and the dependence of this broadening to the variation of coupling strength is quite modest. However, when the collisional dissipation represented by red curves $p_{T}^{\min}=20~$GeV is taken into account , a significantly larger broadening is observed. In this case, the $A_J$ distributions could describe experimental data by ATLAS~\cite{Aad:2010bu} and CMS~\cite{Chatrchyan:2011sx}.  The right panel predicts the sensitivity of $z$ distributions in central Au+Au collisions at $\sqrt{s}=200~$GeV with the coupling strength $g_{med}=2.0$. The QGP medium effects shift the $z$ distributions towards $z=0$ and decrease the averaged $\langle z \rangle$. The strong dependence of imbalance distributions on jet size becomes weaker when collisional dissipation is included~\cite{He2012}.



{\bf Acknowledgments:} This research is supported by the US
Department of Energy, Office of Science,
and by the MOE of China with the Program NCET-09-0411,
by NSF of China with Project Nos. 11075062 and 11221504, 
and in party by CCNU self-determined fundings.

\end{document}